\documentclass[12pt,a4paper,oneside]{article}
\usepackage{graphicx}
\usepackage{overcite}

\title{The Delicate Balance of Static and Dynamic Electron Correlation}
\date{May 20, 2016}
\author{\vspace{1cm}Christopher J. Stein, Vera von Burg and Markus Reiher\footnote{Corresponding author: markus.reiher@phys.chem.ethz.ch}\\
\textit{ETH Z\"urich, Laboratorium f\"ur Physikalische Chemie,}\\ 
\textit{Vladimir-Prelog-Weg 2, 8093 Z\"urich, Switzerland}}

\begin{document}

\maketitle

\begin{abstract}
Multi-configurational approaches yield universal wave function parameterizations that can qualitatively well describe electronic structures along reaction pathways.
For quantitative results, multi-reference perturbation theory is required to capture dynamic electron correlation from the otherwise neglected virtual orbitals.
Still, the overall accuracy suffers from the finite size and choice of the active orbital space and peculiarities of the perturbation theory.
Fortunately, the electronic wave functions at equilibrium structures of reactants and products can often be well described by single-reference methods and hence are accessible to accurate coupled cluster calculations.
Here, we calculate the heterolytic double dissociation energy of four $3d$-metallocenes with the complete active space self-consistent field method and compare to highly accurate coupled cluster data.
Our coupled cluster data are well within the experimental error bars.
This accuracy can also be approached by complete active space calculations with an orbital selection based on information entropy measures.
The entropy based active space selection is discussed in detail.
We find a very subtle balance between static and dynamic electron correlation effects that emphasizes the need for algorithmic active space selection and that differs significantly from restricted active space results for identical active spaces reported in the literature.
\end{abstract}

\newpage

\section{Introduction}
Virtually all wave function based quantum chemical methods construct many-particle basis functions from a single-particle (orbital) basis.
The exponential ansatz of excitation operators in coupled cluster theory leads to the most efficient parameterization of the electronic wave function.
This property constitutes the success of single-reference coupled cluster methods such as coupled cluster with singles, doubles, and perturbatively treated triples excitations (CCSD(T)) in single-configurational cases.
Apart from disconnected excitations introduced by the exponential ansatz, the quantitative accuracy of these methods is rooted in the inherent inclusion of (up to all) virtual orbitals.

For general multi-configurational cases, generalizations to multi-reference coupled cluster are still under development (see, for instance, Refs.\ \citenum{masi98,maha99,hanr05,hana11,evan11}) and have not reached the widespread application of their single-reference counterparts.
As a consequence, complete active space (CAS) approaches with subsequent multi-reference perturbation theory are the method of choice.
All these multi-reference methods introduce a separation of the electron correlation problem, mainly into static and dynamic correlation.
Whereas the orbital selection for the active space from which the multi-reference wave function is constructed considers near-degeneracy of one-particle basis states and hence static correlation effects, the dynamic correlation encoded in the neglected virtual space must be captured in a second step.

A standard approach is to calculate a multi-configurational reference wave function with the complete active space self-consistent field (CASSCF) method\cite{roos80,rued82,wern85,know85} and treat the dynamic correlation by multi-reference perturbation theory (e.g. CASPT2).\cite{ande92}
In CASSCF, the orbital space is divided into three subspaces:
An inactive space where all orbitals are doubly occupied, a virtual space where all orbitals are unoccupied, and an active orbital space within which all possible occupations are considered.
The selection of orbitals to be included in the active space can be based on empirical rules \cite{roos87,pier01,pier03,very11}, natural orbital occupation numbers \cite{jens88,pula88,wout14,pula15,krau14}, or orbital entanglement measures \cite{lege03,riss06,bogu12} as described in our recently proposed automatic selection protocol.\cite{stei16}

Other approaches further subdivide the active space and restrict the number of excitations between the subspaces to allow for larger active spaces than those that can be handled in traditional CASSCF.
Examples are the restricted active space SCF (RASSCF)\cite{olse88,malm90} or generalized active space SCF (GASSCF) methods.\cite{flei01,ma11}

In this work, we investigate how well multi-configurational methods (and in particular the combination CASSCF/CASPT2) are suited to reproduce the energy for complete heterolytic dissociation, called double dissociation in the following, of four metallocenes,
\begin{eqnarray}
\rm M(Cp)_2 \rightarrow  M^{2+} + 2 \, Cp^- .
\nonumber
\end{eqnarray}
Here, M = V, Mn, Fe, and Ni and Cp = cyclopentadiene. For this double dissociation energy, 
theoretical results were published with both single-configurational  \cite{koch96,klop96} and multi-configurational methods \cite{pier95,vanc11,phun12} and also experimental data are available.\cite{ryan92}
We assess the quality of the multi-configurational methods by comparison to experimental results and to highly accurate, explicitly correlated single-reference coupled cluster data.
We further analyze the requirements on the reference wave function that ensures the right balance between static and dynamic correlation and emphasize the role of the active orbital space selection.

\section{Computational Details}

All structures were optimized with Kohn-Sham density functional theory (DFT) employing the PBE0 density functional \cite{perd92,perd96,perd96a} in \textsc{Turbomole} (v.6.5).\cite{ahlr89}
The def2-QZVPP \cite{weig03} and def2-TZVP \cite{weig05} basis sets were chosen for the metal and light atoms, respectively.
Structures obtained with this computational setup were shown to be close to the experimental structures and allow us to compare our calculations with those of Phung \textit{et al.} who employed the same density functional and basis sets.\cite{phun12}
The (heterolytic) double dissociation enthalpy $\Delta H_{298 K}^\circ$ was obtained in a composite approach based on either multi-configurational calculations or single-configurational coupled cluster calculations.
In the multi-configurational approach, the composite scheme consists of four terms:
\begin{equation}
\Delta H_{298 K}^{\circ,\mathrm{CASPT2}} = \Delta E_\mathrm{el}^\mathrm{CASPT2} + \Delta E_\mathrm{CP} + \Delta E_\mathrm{ZPE} + \Delta E_\mathrm{thermal} \, ,
\label{comp_multi}
\end{equation}
where 
\begin{equation}
\Delta E_\mathrm{el}^\mathrm{CASPT2} = E_\mathrm{M^{2+}}^\mathrm{CASPT2} + 2 E_\mathrm{Cp^-}^\mathrm{CASPT2} - E_\mathrm{M(Cp)_2}^\mathrm{CASPT2}  
\end{equation}
is the electronic double dissociation energy obtained by CASPT2, $\Delta E_\mathrm{CP}$ denotes the counterpoise correction according to Boys and Bernardi \cite{boys70}, $\Delta E_\mathrm{ZPE}$ is the zero-point vibrational energy correction and $\Delta E_\mathrm{thermal}$ is the thermal correction calculated for a temperature of 298~K.

We calculated $\Delta E_\mathrm{el}^\mathrm{CASPT2}$ with the IPEA shift set to the recommended value of 0.25~a.u. \cite{ghig04} unless otherwise noted and an imaginary shift \cite{fors97} of 0.1~a.u. was applied to avoid the occurrence of intruder states.
One-electron integrals were calculated for the scalar-relativistic Douglas-Kroll-Hess (DKH) Hamiltonian of second-order in the electron-nucleus potential (DKH2).\cite{hess86,reih04,reih04a}
Two-electron integrals were calculated exploiting Cholesky decomposition.\cite{aqui08}
The ANO-RCC basis set with the [10s\-9p\-8d\-6f\-4g\-2h] contraction for the metal atoms \cite{roos05}, [8s7p4d3f1g] for carbon \cite{roos04} 
(note that we employed the recently corrected carbon basis as implemented in \textsc{Molcas} 8.2) and [6s4p3d1f] for hydrogen \cite{widm90} was applied.
Corresponding active spaces were either selected by our recently proposed automatic selection scheme or were adopted from the RASSCF/RASPT2 calculations of Phung \textit{et al}.\cite{phun12}
The active spaces of the CASSCF calculations are specified by the notation CAS($N,L$)SCF with $N$ being the number of electrons and $L$ the number of spatial orbitals in the active space.

All CASSCF and CASPT2 calculations were calculated with a developers' version of \textsc{Molcas} 8.\cite{aqui8}
The last two contributions in Eq.~(\ref{comp_multi}) were calculated with the \textsc{SNF} \cite{neug02} module from the \textsc{MoViPac} package \cite{weym12} and the necessary single-point calculations were performed with the same DFT setup as for the structure optimization.
The thermal contribution $ \Delta E_\mathrm{thermal}$ to the heterolytic dissociation enthalpy was calculated from standard statistical mechanics relations:
\begin{equation}
  \Delta E_\mathrm{thermal} = \frac{13}{2} RT + \Delta E_\mathrm{vib} ,
\end{equation}
with the ideal gas constant $R$,  temperature $T$=298~K, and the difference in vibrational energies between products and reactants $\Delta E_\mathrm{vib}$.
The factor 13/2 results from considering 1/2 $RT$ per translational and rotational degree of freedom (taking into account that the atomic 
ion can translate, but not rotate) and the relation of enthalpy $H$ and internal energy $U$, $H=U+RT$, for each reactant.

We calculated benchmark data for the heterolytic double dissociation energy of this set of metallocenes based on explicitly correlated coupled cluster theory following a recently proposed protocol for thermochemical calculations for transition metal containing molecules.\cite{bros13}
The corresponding composite scheme is the same as for the multi-reference calculations in Eq.\ (\ref{comp_multi}) 
\begin{equation}
\Delta H_{298 K}^{\circ,\mathrm{CC}} = \Delta E_\mathrm{el}^\mathrm{CC} + \Delta E_\mathrm{CP} + \Delta E_\mathrm{ZPE} + \Delta E_\mathrm{thermal} + \Delta E_\mathrm{DKH} + \Delta E_\mathrm{SC}\, ,
\label{comp_multi2}
\end{equation}
augmented by additional terms that take into account an explicit relativistic correction ($\Delta E_\mathrm{DKH}$) (in our CASPT2 calculations
this is incorporated implicitly) and the semi-core correlation ($\Delta E_\mathrm{SC}$).
Both $\Delta E_\mathrm{ZPE} $ and $ \Delta E_\mathrm{thermal}$ were taken to be the same as for the multi-configurational calculations.
Restricted open-shell explicitly correlated coupled cluster with singles, doubles, and perturbatively treated triples (CCSD(T)-F12b) \cite{know93,know00,adle07,kniz09} was applied for the calculation of the basic contribution $\Delta E_\mathrm{el}^\mathrm{CC}$.
The 'b' denotes an efficient CCSD(T)-F12 approximation.\cite{adle07}

For the coupled cluster calculations, we chose the aug-cc-pVTZ basis set\cite{dunn89,kend92,bala05} as atomic orbital basis and a geminal exponent of $\beta = 1.0 \, \mathrm{a.u.}^{-1}$.
For the density fitting of the Fock and exchange matrices the def2-QZVPP/JKFIT \cite{weig08} and cc-pVTZ/JKFIT \cite{weig02} basis sets for the metal and non-metal atoms, respectively, were chosen.
The aug-cc-pVTZ/MP2FIT basis set \cite{weig02} served for the density fitting of the remaining two-electron integrals for all atoms and as complimentary auxiliary basis set for the resolution of the identity for the metal atoms.
For the non-metal atoms the aug-cc-pVTZ/OPTRI basis set \cite{yous09} was applied for the resolution of the identity.
$\Delta E_\mathrm{DKH}$ and $\Delta E_\mathrm{SC}$ were calculated with conventional CCSD(T).
The former correction was calculated as the difference between calculations with the DKH2 Hamiltonian and the recontracted basis set aug-cc-pVTZ-DK \cite{dejo01} and a calculation without the DKH2 Hamiltonian and the corresponding standard basis set.
The effect $\Delta E_\mathrm{SC}$ of the semi-core correlation from the $3s$- and $3p$-orbitals of the metal atom was calculated as the difference between results for which these orbitals were correlated and those where their occupation was enforced to be doubly occupied in a aug-cc-pwCVTZ basis set.\cite{bala05,pete02}
All coupled cluster calculations were performed with \textsc{Molpro 2010}.\cite{wern12}

Density matrix renormalization group (DMRG) calculations were carried out with \textsc{QCMaquis}.\cite{kell14,dolf14,kell15,kell16}
We adopt the notation 'DMRG[$m$]($N,L$)\-\#\textit{orbital\_basis}', where $m$ is the number of renormalized subsystem states, $N$ and $L$ are the number of active electrons and orbitals, respectively, as before,  and the string after '\#' denotes the orbital basis of the calculation.
We kept 500 renormalized states and applied 20 sweeps in all calculations unless otherwise noted because preliminary calculations showed that the entanglement information is converged with these settings.
A random number guess was applied for the environment in the warm-up sweep because for this no knowledge about the Hartree-Fock determinant is required and convergence to the correct ground state was still observed.
The single-orbital entropy $s_i(1)$, 
\begin{equation}
s_i(1) = - \sum_{\alpha =1}^4 \omega_{\alpha,i} \ln \omega_{\alpha,i} ,
\label{singorb}
\end{equation}
and the mutual information $I_{ij}$  \cite{lege03,riss06,lege06,bogu13} of two orbitals $i$ and $j$ were evaluated from these calculations. 
\begin{equation}
I_{ij} = \frac{1}{2} \left[ s_i(1) + s_j(1) -s_{ij}(2)\right](1-\delta_{ij}) ,
\label{mutinf}
\end{equation}
In Eq.\ (\ref{singorb}), $\alpha$ runs over the four possible occupations in a spatial orbital basis (doubly occupied, spin up, spin down, unoccupied) and $\omega_{\alpha,i}$ are the eigenvalues of the one-orbital reduced density matrix for the $i$-th orbital.
The mutual information requires the two-orbital entropy $s_{ij}(2)$, 
\begin{equation}
s_{ij}(2) = - \sum_{\alpha =1}^{16} \omega_{\alpha,ij} \ln \omega_{\alpha,ij} ,
\end{equation}
calculated from the eigenvalues of the two-orbital reduced density matrix $\omega_{\alpha,ij}$.
The prefactor of 1/2 in Eq.~(\ref{mutinf}) is in accord with the earlier literature,\cite{riss06,bogu13} 
although it has been omitted in most recent work by Legeza and co-workers.\cite{barcz15}
We keep this prefactor in order to be consistent with our recent work.
Note that our implementation of these entropy measures in \textsc{QCMaquis} is not affected by typos that appeared
in Table 3 of Ref.\ \citenum{bogu13} (see Ref.\ \citenum{ma16} for further details).

\section{Results and Discussion}

\subsection{Experimental data and single-configurational calculations}

\begin{figure}[h!]
\includegraphics[width=\textwidth]{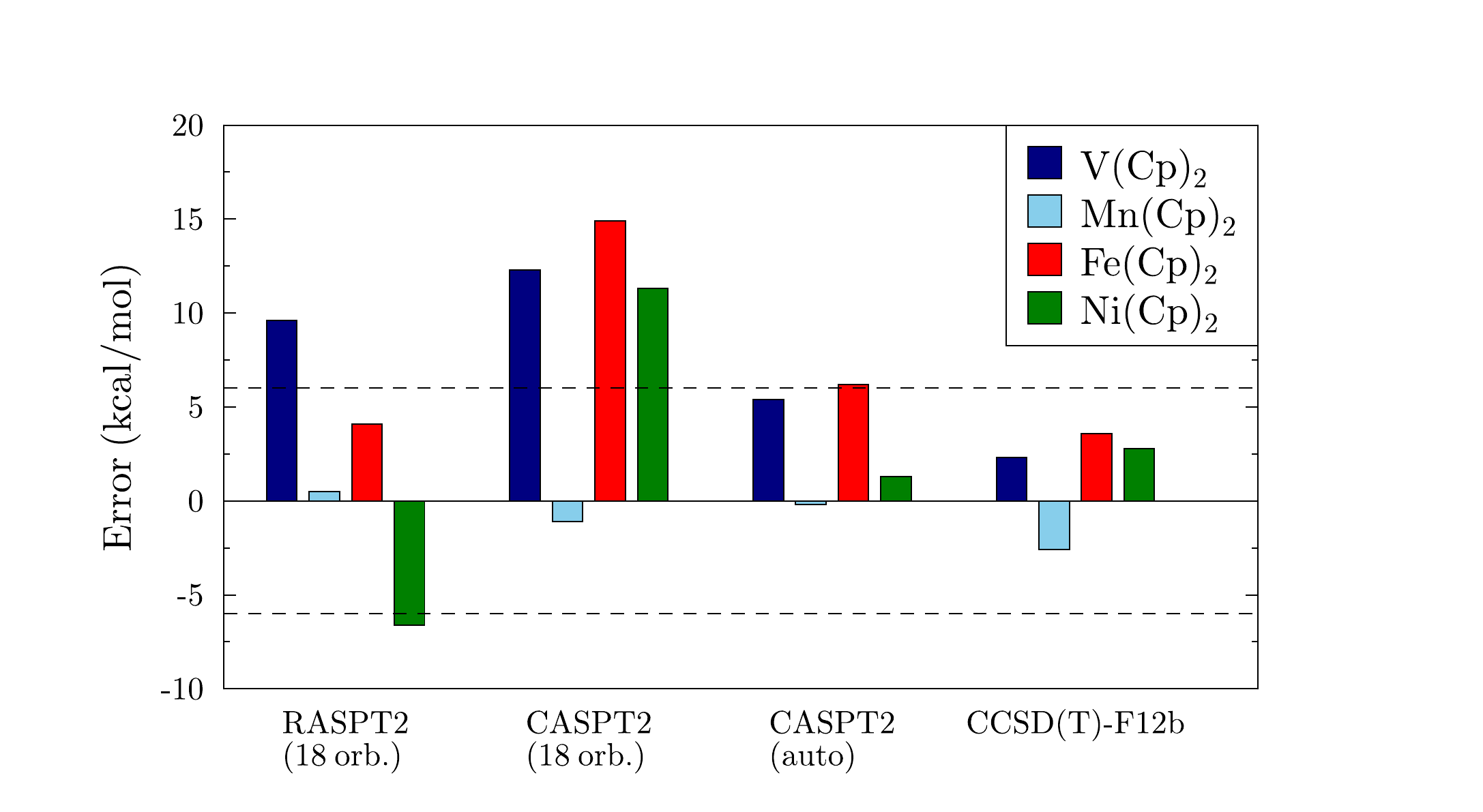}
\caption{Errors of the calculated heterolytic double dissociation energies of metallocenes with respect to experimental data collected in Table \ref{experiment}. The 'RASPT2 (18 orb.)' results were taken from Ref.\ \citenum{phun12}, whereas the 'CASPT2 (18 orb.)' results were obtained for the full active space from the RASPT2 calxulation but in a CASSCF/CASPT2 approach. The 'CASPT2 (auto)' results were obtained for an active space automatically selected based on single-orbital entropies (see text for further details). Dashed horizontal lines indicate the experimental error bars.}
\label{result_bars}
\end{figure}

The experimental heterolytic double dissociation enthalpy $\Delta H_{298 K}^{\circ , \mathrm{exp.}}$ in Table \ref{experiment} is calculated from the standard heats of formation $\Delta H_f^\circ$ of the fragments according to
\begin{equation}
   \Delta H_{298 K}^{\circ,\mathrm{exp.}} = 2 \Delta H_f^\circ \mathrm{(Cp}^-\mathrm{, g)} +  \Delta H_f^\circ \mathrm{(M}^{2+} \mathrm{, g)} - \Delta H_f^\circ \mathrm{(M(Cp)}_2\mathrm{, g)} \, ,
\end{equation}
where 'g' denotes gas-phase measurements.
For the experimental value, the so-called ion convention was chosen meaning that the integrated heat capacity of the electron was assumed to be zero.\cite{lias88}
Furthermore, the adiabatic ionization energy was assumed to be equal to the enthalpy of ionization at 298~K.
We follow Refs.\ \citenum{ryan92, klop96,phun12} and assign experimental error bars of $\pm 6$~kcal/mol in Table \ref{experiment} and Fig.\ \ref{result_bars} although the error bars from the individual contributions would sum to a larger error mostly due to the large uncertainty for  $\Delta H_f^\circ \mathrm{(Cp}^-\mathrm{, g)}$.

\begin{table}[h!]
\caption{Symmetry of electronic states for all complexes and fragments along with their standard heats of formation $\Delta H_f^\circ$ (in kcal/mol) taken from Refs.\ \citenum{crc16,lias88} and resulting heterolytic double dissociation enthalpies $\Delta H_{298 K}^\circ$ in kcal/mol for the reaction M(Cp)$_2 \rightarrow $ M$^{2+} + 2 \, $Cp$^-$ with M = V, Mn, Fe and Ni, Cp = cyclopentadiene. Error bars are given as in Ref.\ \citenum{ryan92}. $\Delta H_f^\circ$(Cp$^-$) amounts to $19.6 \pm 4$~kcal/mol.}
\begin{center}
\begin{tabular}{llllll}
\hline \hline
 & V & Mn & Fe & Ni \\
 M$^{2+}$ &$^4$F (Ref.\ \citenum{suga78}) & $^6$S (Ref.\ \citenum{corl77}) & $^5$D (Ref.\ \citenum{corl82})& $^3$F (Ref.\ \citenum{corl81})\\
 $\Delta H_f^\circ$  & $616\pm2$ & 599 & 655 & 698 \\
 \hline
  M(Cp)$_2$ & $^4$A$_2^{'}$ & $^6$A$_1^{'}$ & $^1$A$_1^{'}$ & $^3$A$_2^{'}$ \\
$\Delta H_f^\circ$  &$49\pm2$ & $66$ & $58 \pm 1$ & $85 \pm 1$\\
  \hline
$\Delta H_{298 K}^{\circ, \mathrm{exp.}}$ & $606\pm6$ & $572\pm6$ & $636\pm6$ & $652\pm6$\\
\hline \hline
\end{tabular}
\label{experiment}
\end{center}
\end{table}

The largest weight, i.e., the largest squared expansion coefficient of a configuration state function ($\approx 0.9$) in the CASSCF calculations and the T1-diagnostic of the coupled cluster calculations \cite{lee89} (lower than 0.02) classify the metallocenes investigated here as single-reference cases.
We therefore exploit high-level coupled cluster data for two main reasons:
(i) to analyze the deviations from experimental data observed for previous theoretical results \cite{klop96} and (ii) to generate theoretical benchmark data for the multi-configurational calculations.

The results in the upper part of Table \ref{result_tab} and Fig.\ \ref{result_bars} show that excellent agreement with experiment can be achieved.
For all metallocenes, the calculated dissociation energies are well within the experimental error bars.
Smaller contributions, however, have to be included because they amount to a total of up to 11~kcal/mol in the case of vanadocene.
The semi-core correlation $\Delta E_\mathrm{SC}$ is approximately proportional to the number of determinants that can be generated with excitations of the semi-core electrons to the valence orbitals and the effect is therefore largest for vanadocene with its $d^3$-configuration.
The counterpoise correction $\Delta E_\mathrm{CPC}$ is necessarily negative for dissociation reactions.
In a dissociation process also the contribution of the zero-point energy $\Delta E_\mathrm{ZPE}$ is negative (the only reported exception being the highly anharmonic ClHCl$^-$ molecule \cite{stei15}) and clearly non-negligible in this case.

Previous coupled cluster calculations yielded a double dissociation energy that is almost 20~kcal/mol higher than the experimental value.\cite{koch96,klop96}
Their smaller corrections agree well with our results so that the deviation can be attributed to either the electronic double dissociation energy $\Delta E_\mathrm{el}^\mathrm{CC}$ or an overestimation of the semi-core correlation.
The latter option is supported by the fact that the authors of Ref.\ \citenum{klop96} estimate the semi-core correlation to amount to 30~kcal/mol based on second-order M{\o}ller-Plesset perturbation theory which is exactly 20~kcal/mol larger than in our calculations.
We assume our explicitly correlated results to be close to the basis set limit, which is supported by the rather small counterpoise corrections.

\begin{table}[ht!]
\caption{Heterolytic double dissociation energies $\Delta H_{298 K}^\circ$ (in kcal/mol) for the reaction
M(Cp)$_2 \rightarrow$ 2 Cp$^- + $ M$^{2+}$ (M = V, Mn, Fe, and Ni) from CCSD(T)-F12b and CASPT2 calculations including correction terms.}
\begin{center}
\begin{tabular}{lrrrr}
\hline \hline
contribution & V(Cp)$_2$ & Mn(Cp)$_2$ & Fe(Cp)$_2$ & Ni(Cp)$_2$ \\ 
\hline
$\Delta E_\mathrm{CCSD(T)}$ &597.1 & 564.9 & 629.8 & 653.2   \\
$\Delta E_\mathrm{SC}$  & 15.5 & 7.9 & 10.7 & 3.5 \\
$\Delta E_\mathrm{DKH}$  & 3.1 & 2.8 & 6.6& 5.0 \\
$\Delta E_\mathrm{CPC}$  & $-1.4$ & $-1.1$ & $-1.8$ & $-1.4$ \\
$\Delta E_\mathrm{ZPE}$  & $-7.3$ &$-5.5$ &$-8.7$ & $-6.6$\\
$\Delta E_\mathrm{thermal}$  & 1.4 & 0.5 & 1.9 & 1.0 \\
$\Delta H_{298}^0$  (CC)& 608.3 & 569.4 & 638.6 & 654.8 \\
\hline
$\Delta E_\mathrm{CASPT2}$ & 622.6 &578.6 & 650.6 & 663.9 \\
$\Delta E_\mathrm{CPC}$  & $-5.3$ & $-1.8$ & $-2.7$ & $-5.0$ \\
$\Delta E_\mathrm{ZPE}$  & $-7.3$ &$-5.5$ &$-8.7$ & $-6.6$\\
$\Delta E_\mathrm{thermal}$  & 1.4 & 0.5 & 1.9 & 1.0 \\
$\Delta H_{298}^0$ (CASPT2)  & 611.4 & 571.8 & 641.2 & 653.3 \\
\hline
exp. &$606 \pm 6$ & $572 \pm 6 $ & $635 \pm 6$ & $652 \pm 6$ \\
\hline \hline
\end{tabular}
\label{result_tab}
\end{center}
\end{table}

\subsection{Multi-configurational calculations: the fragments}

The calculation of CASPT2 energies for the cyclopentadiene ion included six electrons in five $\pi$-orbitals in the active space.
That is one bonding orbital more compared to the calculation of Phung \textit{et al.}\cite{phun12}, but the effect on the energy turned out
to be less than 0.5~kcal/mol.

Considering that small transition metal compounds with free valences are particularly challenging for electronic structure calculations,
the metal cations in this study are worst cases as all valences are unsaturated.
We investigated several compositions of active spaces for the bare metal cations and found an increasing spread of the CASPT2 energy
obtained for different active-space selections with increasing total electron number (see Table \ref{cations}).
The smallest active space consisted of the $3d$-orbitals only and is our reference for Table \ref{cations}.
A second set of $d$-orbitals ($3d'$) is included in the first row of that table to account for the double-shell effect.\cite{ande95}
This active space was chosen for all cations but V$^{2+}$ in the study by Phung \textit{et al.}\cite{phun12} where the $3d'$-orbitals formed the RAS3 space in their RASSCF calculations.
The inclusion of the second $d$-shell has a dramatic effect on the energy, especially in the case of Ni$^{2+}$.
An increasing change of the energy due to the double-shell effect is expected with increasing number of $d$-electrons \cite{pier01} and observed here as well. 

\begin{table}[h!]
\caption{CASPT2 energy differences (in kcal/mol) for different CAS sizes relative to a CAS including only the $3d$-orbitals (5 orb.) for the atomic dications.
The last row lists CASPT2 energy differences obtained for a CAS that was automatically selected based on single-orbital entropies and the
number of selected orbitals for each atomic dication is given in parentheses.}
\begin{center}
\hspace*{-1cm}
\begin{tabular}{lcccc}
\hline \hline
active orbitals & V$^{2+}$ & Mn$^{2+}$ & Fe$^{2+}$ & Ni$^{2+}$ \\ 
\hline
$3d, 3d'$ (10 orb.)            & 1.31  & 8.95  & 14.30 & 28.38 \\
$3d, 4s, 4p$ (9 orb.)         & -0.52 & -0.27 & 3.44   & -0.82    \\
$3d, 3d', 4s, 4p$ (14 orb.)& 1.29  & 9.73  & 15.30 & 26.53   \\
\hline
automated selection          & -2.69 (11 orb.) & 2.47 (8 orb.)  & 2.40 (8. orb)  & 29.77 (15 orb.)   \\
\hline \hline
\end{tabular}
\label{cations}
\end{center}
\end{table}

A full-valence active space including the unoccupied $4s$- and $4p$-orbitals in addition to the $3d$-orbitals introduces only minor changes to the energy amounting to at most 3.44~kcal/mol in the case of ferrocene.
Consequently, an active space consisting of all the valence orbitals and the $3d'$-orbitals (third row in Table \ref{cations}) yields energies very close to those obtained when only the double-shell orbitals are included.

Finally, we applied our recently proposed automated active space selection \cite{stei16} to identify a suitable active space.
Starting from CASSCF orbitals with an active space including only the $3d$-orbitals, we examined the single-orbital entropy of all valence orbitals and of the $3s$- and $3p$-orbitals.

Our automated selection procedure included the $3d$ and the $3p$ semi-core orbitals in the active space for all cations.
In the case of V$^{2+}$, three additional $3d'$-orbitals were selected, but for this cation the energy was the least sensitive to the choice of the CAS.
Ni$^{2+}$ required the largest active space consisting of the $3p$, $3d$, $4s$, $4p$, and three $3d'$-orbitals.
Although only a part of the second $d$-shell is included in this calculation, a comparison with the energy from calculations including all $3d'$-orbitals reveals that the double-shell effect is fully accounted for.
Hence, the selection of a proper active space for these cations is a very delicate task and can introduce an error to the double dissociation energy of up to 30~kcal/mol.

While we found that our automated selection procedure\cite{stei16} works well for these atomic dications, we note that for Ni$^{2+}$ the partially
converged DMRG calculation required a number of renormalized states $m$ as high as 2000 (instead of typically 500) in order to obtain 
qualitatively correct single-orbital entropies. This, however, is easily detected because the preliminary and the converged single-orbital entropies deviate
for smaller $m$ values. In the case of Ni$^{2+}$,
partially converged DMRG calculations were carried out for $m$=500, 1000, and 1500, until the entropies were detected unchanged for $m$=2000.

\subsection{Multi-configurational calculations: ferrocene}

\begin{figure}[h!]
\includegraphics[width=\textwidth]{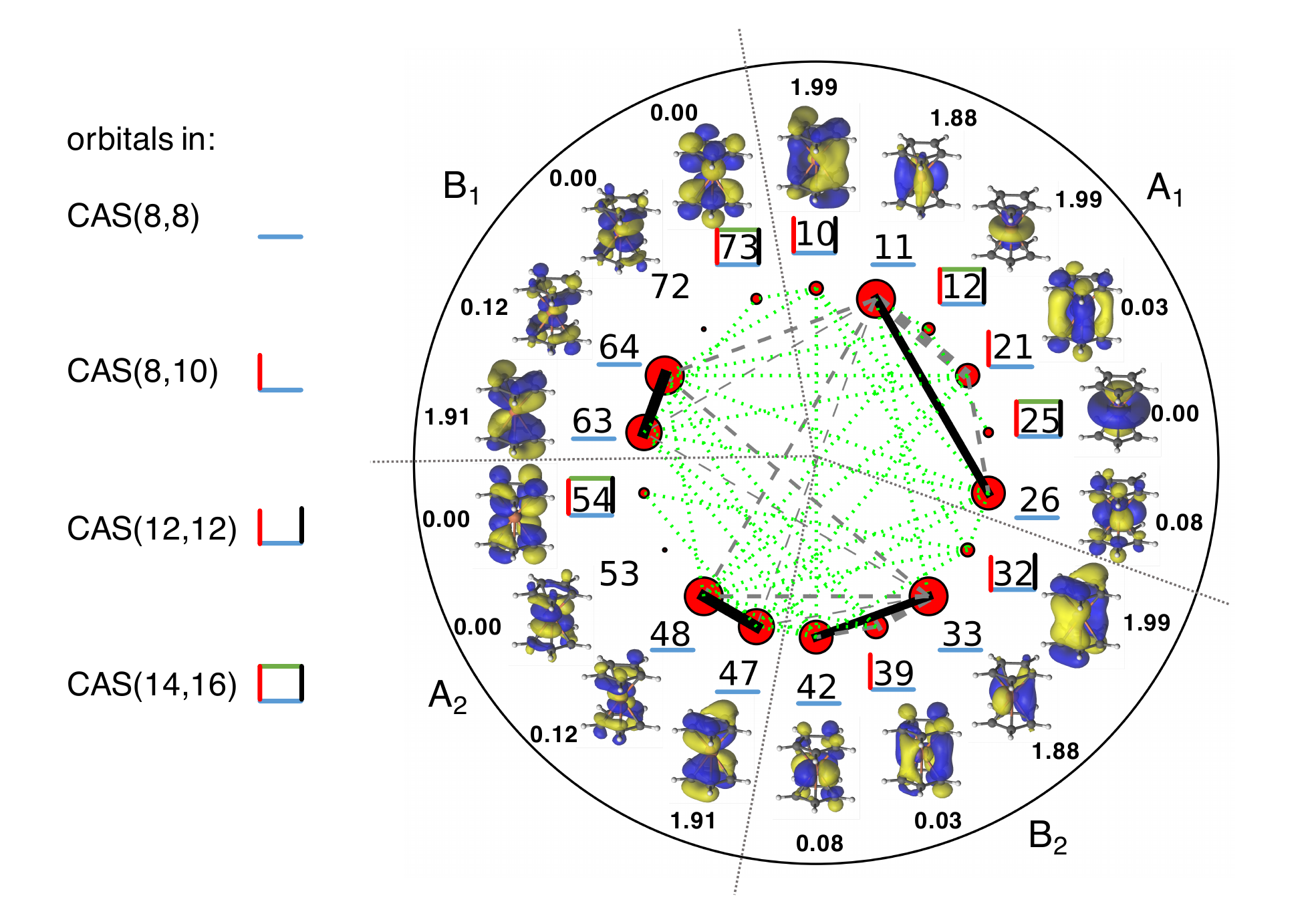}
\caption{Reduced entanglement diagram for ferrocene from a DMRG[500](66,73)\#CAS(6,5)SCF calculation. Out of the 73 spatial orbitals, the diagram shows the 18~orbital active space selected by Phung \textit{et al.}\cite{phun12} along with the orbital number in the DMRG calculation ordered according to the four irreducible representations $A_1$, $A_2$, $B_1$, and $B_2$ in C$_{2v}$ symmetry. The area of the red circles is proportional to an orbital's single-orbital entropy, while the thickness of the connecting lines is proportional to their mutual information (green dashed lines indicate a value~$> 0.001$, gray dashed lines correspond to a value~$ > 0.01$, and values~$>0.1$ are indicated by a solid black line). DMRG natural occupation numbers are given in bold face. The orbital pictures were obtained from the converged CAS(14,18)-SCF calculation including the whole set of orbitals shown here. The active spaces are indicated by the colored bars surrounding the orbital numbers as defined on the left-hand side and the large active spaces include all orbitals of the small active spaces.}
\label{ferrocene}
\end{figure}

We choose ferrocene as a first example for the analysis of a suitable active space and of the effect of the choice of the CAS on the final CASPT2 energies.
In Fig.\ \ref{ferrocene}, an entanglement diagram including all active orbitals picked by Phung \textit{et al.} \cite{phun12} for the RAS1 to RAS3
spaces is presented along with orbital pictures and their natural occupation numbers.
Both, occupation numbers and entanglement information, are obtained from a DMRG[500](66,73) \#CAS(6,5)SCF calculation, where the orbitals were selected such that they span a certain energy range around the Fermi level which includes all double-shell orbitals.
The orbital pictures correspond to optimized orbitals from a CAS(14,18)SCF calculation including all orbitals shown in the diagram.

\begin{figure}[h!]
\includegraphics[width=\textwidth]{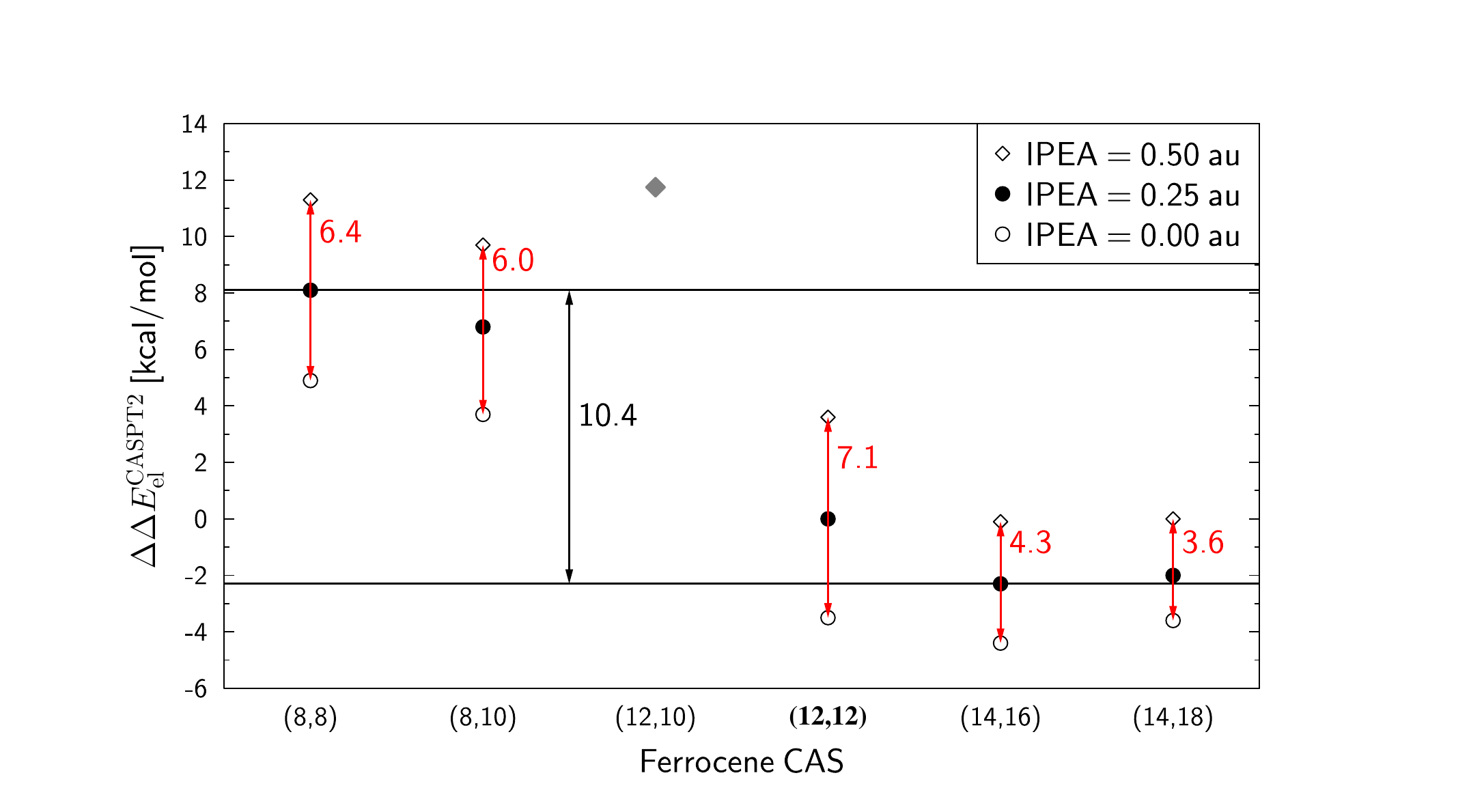}
\caption{CASPT2 electronic double dissociation energy differences $\Delta \Delta E_\mathrm{el}^\mathrm{CASPT2}$ (in kcal/mol) for different active spaces and IPEA shifts relative to the CAS(12,12) active space with the recommended IPEA shift of 0.25~a.u. The solid black horizontal lines highlight the energy range covered by the variation of the active space size for the set of calculations with the recommended IPEA shift of 0.25~a.u. Red numbers denote the energy range covered by the variation of the IPEA shift for a given active space size. The gray diamond indicates the result of a CASPT2 calculation with a CAS(12,10) selected as described in the text.}
\label{ferrocene_energy}
\end{figure}

\begin{figure}[h!]
\hspace*{-0.6cm}\includegraphics[width=1.1\textwidth]{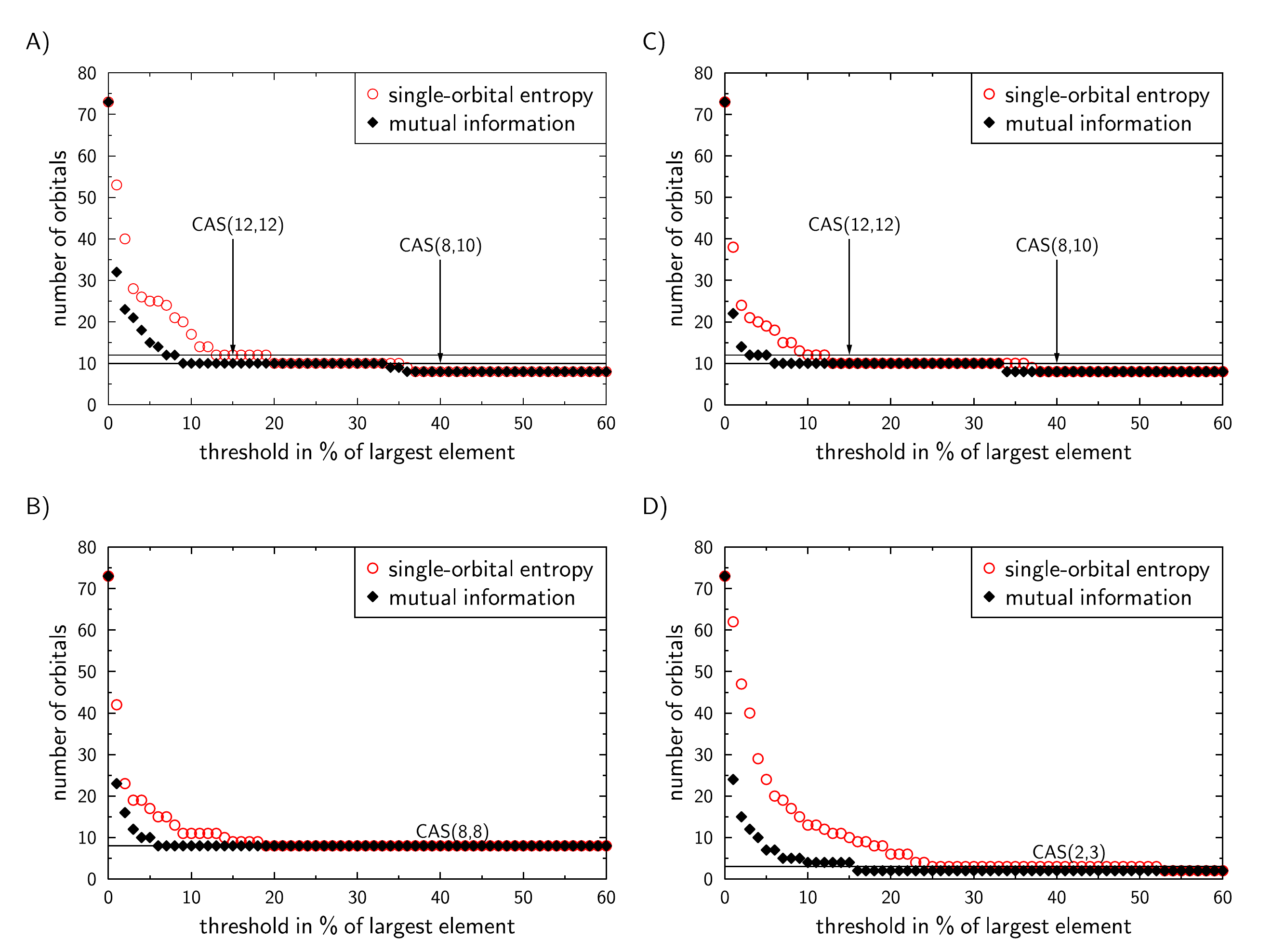}
\caption{Threshold diagrams for the single-orbital entropy (red) and the mutual information (black). The selection threshold is the fraction of the maximum value of the single-orbital entropy or mutual information for each calculation given on the abscissa. The ordinate shows the number of selected orbitals. The four panels correspond to DMRG[500](66,73) calculations with different orbital bases: panel A) CAS(4,4)SCF where the orbitals were selected from around the Fermi level, panel B) CAS(8,8)SCF also with orbitals around the Fermi level, panel C) CAS(6,5)SCF including all $3d$-orbitals and panel D) CAS(6,5)SCF with randomly chosen, spatial-symmetry breaking orbitals.}
\label{orbital_bases}
\end{figure}

To evaluate whether the CASPT2 energies are converged with respect to the active space size, we performed CASPT2 calculations with several active orbital subsets shown in Fig.\ \ref{ferrocene} (left).
The subsets were chosen such that groups of orbitals with similar single-orbital entropy were subsequently added to a minimal CAS consisting of eight orbitals.
Fig.\ \ref{ferrocene_energy} shows the electronic double dissociation energy differences $\Delta \Delta E_\mathrm{el}^\mathrm{CASPT2}$ obtained from CASPT2 calculations with varying active space size.
The CAS(12,12) results were chosen as the reference because this CAS was automatically selected based on entropy selection criteria as will be discussed below.
The minimal CAS(8,8) active space consists of $3d_{xy}$- and $3d_{x^2-y^2}$-orbitals and corresponding double-shell orbitals and a set of degenerate pairs of $\pi$-  and $\pi^*$-orbitals.
Two additional non-bonding orbitals are added to form a CAS(8,10) but the effect on $\Delta E_\mathrm{el}^\mathrm{CASPT2}$ is only about 1~kcal/mol.
The addition of two bonding Cp$^-$ $\pi$-orbitals, however, leads to a significant energy decrease of almost 7~kcal/mol.
Inclusion of more orbitals leaves the double dissociation energy $\Delta E_\mathrm{el}^\mathrm{CASPT2}$ almost unchanged so that we conclude that the same accuracy as for the large CAS(14,18) can be obtained with the CAS(12,12).
This CAS(12,12) gives rise to a plateau in the orbital discarding threshold diagrams \cite{stei16} (cf.\ Fig.\ \ref{orbital_bases}, panels A and C) and will therefore be the active space picked by the automated selection algorithm as discussed below.
The energy decrease from the CAS(8,8) to the CAS(8,10) calculation is small compared to the inclusion of two additional orbitals in CAS(12,12).
A calculation with orbitals 10 and 32 in Fig.\ \ref{ferrocene} added to the CAS(8,8) shows that the two initially added orbitals (no. 21 and 39 in Fig.~\ref{ferrocene}) are essential.
The corresponding energy difference $\Delta \Delta E_\mathrm{el}^\mathrm{CASPT2}$ is included as a gray diamond in Fig.\ \ref{ferrocene_energy}.
The raise in energy indicates an imbalanced CAS and shows that the selection based on single-orbital entropies 
(and in this case also the natural occupation numbers) lead to a monotonically convergent behavior of the energies upon inclusion of more active orbitals.

As can be seen from Fig.\ \ref{ferrocene}, an occupation number based orbital selection will yield a CAS(8,10) if orbitals with occupation numbers between 0.02 and 1.98 are included.
The two additional orbitals in the CAS(12,12) would not be selected and an error of 6~kcal/mol would remain.
Certainly, this can in principle be overcome by extending the range of occupation numbers for orbitals to be included in the active space.
However, many orbitals may have similar occupation numbers especially for large molecules.
A slight extension of the occupation number range will therefore result in the inclusion of many more orbitals and a clear-cut
enlarged interval will be difficult to define unambiguously. By contrast, our
scaling of the selection threshold to the maximum element of the single-orbital entropy in our automated approach induces a large spread of the quantity on which the selection is based.
In this way, our selection criterion adapts to the static correlation present in the molecule under study.
Furthermore, our selection threshold is not a fixed fraction of the maximum single-orbital entropy but is flexibly adapted to identify groups of orbitals with similar single-orbital entropy that can be identified as plateaus in the threshold diagrams.
This further adapts the orbital selection procedure to the molecule under study and allows an automated selection even for large molecules.

A peculiarity of CASPT2, the IPEA shift is an empirical parameter with a recommended value of 0.25~a.u.\ derived from a limited test set of mostly diatomic molecules with rather small active spaces.\cite{ghig04}
Naturally, it was suggested to either increase the IPEA shift to a value of 0.5-0.7~a.u.\ \cite{daku12,kepe09}, one of less than 0.2~a.u.\ \cite{quer08}, or to neglect it completely \cite{frac15} because no effect was observed for the systems studied.
Fig.\ \ref{ferrocene_energy} also includes the dissociation energies for the various CAS sizes from calculations employing an IPEA shift of 0.0 and 0.5~a.u. for both reactants and products.
Importantly, we find that the energy range generated by varying this empirical parameter is larger than the energy difference between the CAS(12,12) and CAS(14,16) calculations.
It therefore defines a lower limit to the accuracy that can be obtained.
For future applications, it might be of advantageous to re-evaluate the IPEA shift with an extended test set (although this would render the sophisticated CASPT2 approach somewhat semi-empirical) or apply parameter-free definitions of the zeroth-order Hamiltonian as in $n$-electron valence state perturbation theory (NEVPT2).\cite{ange01,ange01a,knec16}

The automated selection of the active orbital space requires an orbital basis for the initial DMRG calculation.
In Ref.\ \citenum{stei16}, we recommended CASSCF orbitals with a minimal active space as a suitable orbital basis.
The choice of this active space, however, is not unique as it is determined by the desired accuracy achieved for a specific
electronic structure model (CASSCF in Ref.\ \citenum{stei16}). We study here the automated selection for several choices of CASSCF orbital bases
in CASPT2 calculations.

Fig.\ \ref{orbital_bases} collects threshold diagrams from partially converged DMRG calculations for four different orbital bases.
The threshold diagrams show the number of orbitals whose single-orbital entropy (or largest mutual-information element) lies above a threshold defined as a fraction of the largest single-orbital entropy (or mutual information element) of a given calculation.
Plateaus in these diagrams indicate groups of orbitals with similar entropy.

We find that the first plateau (with the largest number of orbitals) of a given threshold diagram defines a set of orbitals that forms a suitable active space for a CASPT2 calculation.
In panel A, the two degenerate highest occupied molecular orbitals (HOMOs) and the two lowest unoccupied molecular orbitals (LUMOs) form a CAS(4,4) orbital basis for a DMRG[500](66,73)\#CAS(4,4)SCF calculation.
Four additional orbitals (HO\-MO-1 and LUMO+1) are included in panel B.
The metal $3d$-orbitals form the CAS(6,5)SCF orbital basis in panel C and the result of a calculation based on a randomly chosen CAS(6,5) is shown in panel D.
As previously observed \cite{stei16}, we cannot extract additional information for the orbital selection from the mutual information but as it might be useful in a different context, we also report these data.
In panels A and C, a plateau is found that defines a CAS(12,12) and a more pronounced plateau corresponding to a CAS(8,10).

As discussed above, the CAS(12,12) gives a double dissociation energy close to the converged energy while the CAS(8,10) results in an error of about 6~kcal/mol.
However, our previous definition \cite{stei16} of a plateau as a threshold range of at least ten percent has then to be revised to a lower value of 3 to 5~\%
for CASPT2 calculations.
Hence, we understand that it is beneficial for the automated active-orbital selection to automatically analyze the effect of the addition of an orbital (or a pair of degenerate orbitals) to the active space that is ranked next but not included in the set of orbitals creating a plateau in the threshold diagram.

The CAS(12,12) does not give rise to a plateau in panel B of Fig.\ \ref{orbital_bases}.
We conclude that a minimalistic approach of including either only the HOMO and LUMO orbitals or the $d$-orbitals for transition metal systems in an initial CASSCF calculation enables the selection of a suitable active space.
The random selection of initial active orbitals in panel D necessarily fails because the resulting CASSCF orbitals are symmetry broken.
This minimalistic approach for the selection of the initial active space in the preparatory CASSCF calculation therefore requires only information about degenerate orbitals to avoid symmetry breaking.
It can further be assumed that the statically correlated orbitals are close to the Fermi level so that the orbitals for the partially converged DMRG calculation can be chosen automatically from a given energy range around the Fermi level.

\subsection{Multi-configurational calculations: vanadocene, man\-ga\-no\-cene, nickelocene}

\begin{figure}[h!]
\begin{center}
\includegraphics[width=0.6\textwidth]{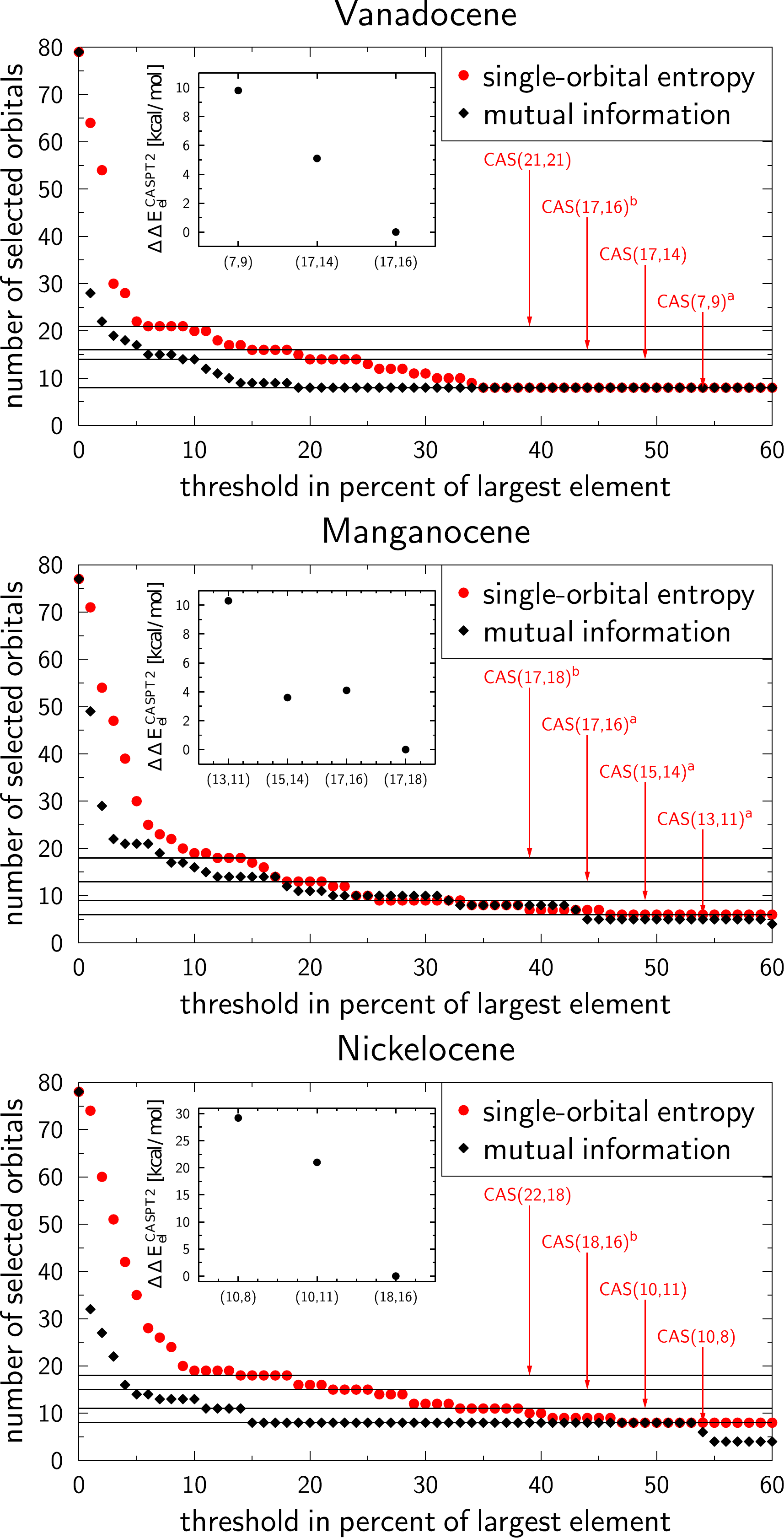}
\caption{Threshold diagrams for vanadocene, manganocene, and nickelocene. Relative double dissociation energies $\Delta \Delta E_\mathrm{el}^\mathrm{CASPT2}$ for the different CAS sizes with respect to double dissociation energy obtained for the largest active space are given as an inlay. A superscript 'a' to the CAS($N$,$L$) notation denotes active spaces that include only one out of two degenerate orbitals in the partially converged calculation and the degenerate orbital was added so that no symmetry-broken solution was obtained. A superscript 'b' denotes the active space selected for the double dissociation energy.}
\label{other_met}
\end{center}
\end{figure}

Threshold diagrams for the remaining metallocenes are shown in Fig.\ \ref{other_met} and the active spaces for several plateaus are specified in red.
The inlays show the energy differences with respect to the largest active space considered for the given molecule.
In case of vanadocene (upper panel in Fig.\ \ref{other_met}), the CAS(21,21) was too large for conventional CASSCF and could not be considered in this study because of the unfeasibility of subsequent standard CASPT2 calculations.
However, the inlay shows that the energy is not converged with respect to the active space yet.
In fact, the addition of the three orbitals in the large CAS(21,21) might then even reduce the error of about 5~kcal/mol in comparison with experiment.

In certain cases, the entanglement information from partially converged DMRG calculations might not result in the same single-orbital entropy for a pair of degenerate orbitals.
We observed this in the calculations on man\-ga\-no\-cene and enforced the addition of energetically degenerate orbitals when one orbital of a pair was selected.
The CASPT2 energy was constant upon enlarging the active space from a CAS(15,14) to a CAS(17,16) but decreased by about 4~kcal/mol for the large CAS(17,18).
This CAS was therefore selected although the energy convergence is not as good as for ferrocene.

For nickelocene, the large CAS(22,18)SCF calculation introduced a slight spatial-symmetry breaking of degenerate orbitals which ultimately led to a decrease in the CASPT2 energy compared to the CAS(18,16) of almost 70~kcal/mol.
This is an artifact that might be avoided by tighter convergence thresholds which, however, would make the calculation too costly.
Although the inlay indicates that convergence with the active space is not achieved, we chose the CAS(18,16) result noting that the correct energy is likely to be lower.

We conclude that quantitative results for the heterolytic dissociation energies of all metallocenes can be obtained if the active space is selected based on orbital entanglement criteria.
Despite the excellent agreement, we do not expect our results to be more accurate than the coupled cluster data.
The accuracy that can be obtained by M{\o}ller-Plesset perturbation theory of second order (the single-reference counterpart of CASPT2) is expected to be lower than CCSD(T) and we therefore do not expect CASPT2 to be significantly more accurate for single-reference cases.
Therefore, we are aware of the fact that the CASPT2 double dissociation energies might benefit from error compensation.
The size of the CAS is certainly the main limitation in the case of vanadocene and nickelocene but it can be overcome by DMRG calculations with a perturbation theory approach for the dynamic correlation.\cite{kura11,kura14,knec16,guo16,soko16,wout16}

\subsection{Comparison of restricted and complete active space approaches}
Phung \textit{et al.} \cite{phun12} reported accurate data for the heterolytic double dissociation energies from RASSCF/RASPT2 calculations.
Since the active spaces chosen in their study involve at most 18 orbitals, we were able to calculate CASPT2 dissociation energies with the same active spaces to quantify the error introduced by the restrictions in the excitation patterns in the RAS1 and RAS3 subspaces.

The two sets of calculations are compared on the left-hand side of Fig.\ \ref{result_bars}.
While the discrepancy is small for V(Cp)$_2$ and Mn(Cp)$_2$, it is significant for the two other metallocenes amounting to almost 20~kcal/mol in the case of Ni(Cp)$_2$.
The reason is rooted in the implementation of the RASSCF/RASPT2 in \textsc{Molcas} as has been discussed in detail in Ref.\ \citenum{vanc11}.
Since orbital rotations between different RAS subspaces are not allowed, the diagonalization of the active part of the Fock matrix is not complete and nonzero elements remain in the parts that couple different RAS subspaces.
These nonzero elements are then simply neglected so that the coupling is ignored in the perturbation treatment.
Consequently, strongly coupled orbitals need to be in the same RAS subspace.

Such strongly coupled orbitals can be identified by their mutual information as displayed in Fig.\ \ref{ferrocene}.
A comparison with the definition of the RAS subspaces in Ref.\ \citenum{phun12} reveals that the $3d'$ double-shell orbitals 26 and 42 are strongly entangled with their corresponding $3d$ counterparts.
However, the former are in RAS3 while the latter are in RAS2 and the corresponding coupling elements are presumably large.
This also explains why the effect is less severe for vanadocene and manganocene because we do not observe the strong entanglement to double-shell orbitals in the metallocenes with less $3d$-electrons.
Hence, the mutual information guides the proper selection of orbitals for the different RAS subspaces whenever a RASSCF/RASPT2 approach is applied.
The good agreement of the RASPT2 dissociation energies with experiment is therefore most likely due to a fortuitous error compensation with the calculation for the metal ions.

\clearpage

\section{Conclusions}
In this work, we calculated the heterolytic double dissociation energy of four $3d$-metallo\-cenes with a highly accurate composite coupled cluster approach and a multi-con\-fig\-urat\-ion\-al approach with perturbation theory.
Although the accuracy of the coupled cluster approach is unrivaled, our multi-configurational calculations give excellent results with a proper selection of the active orbital space.
Both sets of data lie well within the experimental error bars.
This shows that multi-configurational approaches will yield accurate results in the limit of single-reference cases if the balance between static and dynamic correlation is adequately described.

In order to achieve this for CASPT2 calculations, we had to slightly adapt in our automated orbital selection algorithm the definition of a plateau 
from 10~\% to 3~\% in threshold diagrams.
We emphasize that this adjustment does not affect the results or conclusions drawn in Ref.\ \citenum{stei16}.
The metallocenes we studied here are typical single-configurational cases while all molecules in Ref.\ \citenum{stei16} were selected because they were known to be strongly statically correlated.
For single-reference cases, the spread of the single-orbital entropies is certainly smaller than for strongly statically correlated molecules and groups of orbitals with a similar single-orbital entropy are more difficult to identify which is then manifested in less pronounced plateaus in the threshold diagrams.
Minor changes or generalizations of the selection protocol, however, are to be expected when the scope of application of the automated orbital selection is widened.
An analysis of several CASSCF orbital bases for the preparatory DMRG calculation revealed that a minimal approach including only HOMO and LUMO or the metal $d$-orbitals was best suited for the automated selection.

We considered in detail the uncertainty introduced by the semi-empirical IPEA shift in the CASPT2 Hamiltonian.
Future work will reveal whether alternative and parameter-free formulations of a multi-reference perturbation theory such as NEVPT2\cite{ange01,ange01a,knec16} can overcome this problem. 
The RASPT2 data of Ref.\ \citenum{phun12} was reviewed from the point of view of CASPT2 results calculated for the same active spaces.
Significant differences were observed for the $3d$-electron rich metallocenes.
This was rationalized by a combination of the increasingly important double-shell effect and the neglect of coupling elements between different RAS spaces in the Fock matrix of the RASPT2 implementation in \textsc{Molcas}.
The strong correlation between certain metal $3d$-orbitals and their double-shell counterparts results in large coupling elements whose neglect introduces a significant error.
This can be overcome by exploiting entanglement measures when defining the RAS subspaces or abandon the excitation restrictions in the RAS concept by methods such as DMRG or full configuration interaction quantum Monte Carlo.\cite{boot09,shep12}

We showed that quantitative results can be obtained from CAS wave functions with active orbital spaces selected with our entanglement based selection protocol.
In future work, we will further validate the suitability of our automated orbital selection algorithm for excited states and reaction pathways.

\section*{Acknowledgements}
This work was supported by the Schweizerischer Nationalfonds (Project No. 200020\_156598).
CJS thanks the Fonds der Chemischen Industrie for a K\'ekule fellowship.


\providecommand{\url}[1]{\texttt{#1}}
\providecommand{\urlprefix}{}
\providecommand{\foreignlanguage}[2]{#2}
\providecommand{\Capitalize}[1]{\uppercase{#1}}
\providecommand{\capitalize}[1]{\expandafter\Capitalize#1}
\providecommand{\bibliographycite}[1]{\cite{#1}}
\providecommand{\bbland}{and}
\providecommand{\bblchap}{chap.}
\providecommand{\bblchapter}{chapter}
\providecommand{\bbletal}{et~al.}
\providecommand{\bbleditors}{editors}
\providecommand{\bbleds}{eds.}
\providecommand{\bbleditor}{editor}
\providecommand{\bbled}{ed.}
\providecommand{\bbledition}{edition}
\providecommand{\bbledn}{ed.}
\providecommand{\bbleidp}{page}
\providecommand{\bbleidpp}{pages}
\providecommand{\bblerratum}{erratum}
\providecommand{\bblin}{in}
\providecommand{\bblmthesis}{Master's thesis}
\providecommand{\bblno}{no.}
\providecommand{\bblnumber}{number}
\providecommand{\bblof}{of}
\providecommand{\bblpage}{page}
\providecommand{\bblpages}{pages}
\providecommand{\bblp}{p}
\providecommand{\bblphdthesis}{Ph.D. thesis}
\providecommand{\bblpp}{pp}
\providecommand{\bbltechrep}{Tech. Rep.}
\providecommand{\bbltechreport}{Technical Report}
\providecommand{\bblvolume}{volume}
\providecommand{\bblvol}{Vol.}
\providecommand{\bbljan}{January}
\providecommand{\bblfeb}{February}
\providecommand{\bblmar}{March}
\providecommand{\bblapr}{April}
\providecommand{\bblmay}{May}
\providecommand{\bbljun}{June}
\providecommand{\bbljul}{July}
\providecommand{\bblaug}{August}
\providecommand{\bblsep}{September}
\providecommand{\bbloct}{October}
\providecommand{\bblnov}{November}
\providecommand{\bbldec}{December}
\providecommand{\bblfirst}{First}
\providecommand{\bblfirsto}{1st}
\providecommand{\bblsecond}{Second}
\providecommand{\bblsecondo}{2nd}
\providecommand{\bblthird}{Third}
\providecommand{\bblthirdo}{3rd}
\providecommand{\bblfourth}{Fourth}
\providecommand{\bblfourtho}{4th}
\providecommand{\bblfifth}{Fifth}
\providecommand{\bblfiftho}{5th}
\providecommand{\bblst}{st}
\providecommand{\bblnd}{nd}
\providecommand{\bblrd}{rd}
\providecommand{\bblth}{th}

%
%
%

\end{document}